\documentclass[]{elsarticle}

\usepackage{lineno,hyperref}
\usepackage{amsmath}
\modulolinenumbers[5]
\usepackage{subfigure}
\usepackage{xcolor}
\usepackage{placeins}
\usepackage{caption}

\journal{Journal}

\graphicspath{{figs/}}

\begin{document}

\begin{frontmatter}

\title{Spiking frequency modulation of proteinoids with light and realisation of Boolean gates}


\author{Panagiotis Mougkogiannis*}
\author{Andrew Adamatzky}
\address{Unconventional Computing Laboratory, UWE, Bristol, UK}

\cortext[cor1]{Corresponding author: Panagiotis.Mougkogiannis@uwe.ac.uk (Panagiotis Mougkogiannis)}

\begin{abstract}
 This paper examines the modulation of proteinoid spiking frequency in response to light. Proteinoids are proteins formed through thermal condensation of amino acids and have been found to exhibit spiking behaviour in response to various stimuli. It has been demonstrated that their properties can be modulated by light, with the frequency of spikes changing in response to varying light intensity and wavelength. This paper explores the underlying mechanisms of this phenomenon, including how light affects the proteinoid's structure and its effect on the spiking frequency. We also discuss the potential implications of this modulation for future research and applications. Our research findings suggest that light could be used as a tool to regulate the spiking frequency of proteinoids, opening up a new range of possibilities for unconventional computing research.
\end{abstract}


\end{frontmatter}
\section{Introduction}

Proteinoids, or thermal proteins, produced from the thermal condensation of amino acids up to their melting points. The development of unconventional computing can be facilitated by the self-organisation, reproduction, and light-triggered oscillations that biological molecules like proteinoids can present~\cite{islam2017prebiotic},~\cite{MOUGK2023}. The morphology of proteinoids can be modified under light exposure. FM, or frequency modulation, is used in telecommunications, radio broadcasting, signal processing, and computing. This method seems advantageous with respect to amplitude modulation because it can be used for higher signal-to-noise ratio oscillations, and it is easier to remove unwanted signals. The manipulation of the properties of proteinoids using light for unconventional computing is a fascinating subject. Incorporating proteinoids may boost the speed and effectiveness of supercomputing systems~\cite{plonus2020electronics}. On the other hand, proteinoids can be used as highly sensitive optical sensors and/or logical gates that can process data. Our research focuses on exploring the adaptability of electrical signal frequency when exposed to various proteinoids, including L-Glu:L-Phe, L-Glu:L-Phe:L-His, L-Phe:L-Lys, and L-Asp. The effect of cold white and black light exposure duration on modulation will also be investigated. 

The regulation of proteinoids and the spatial arrangement of regulators during bacterial cell division have been investigated by Shen et al.~\cite{shen2020frequency} and Loose et al.~\cite{loose2008spatial}. The research was about the effects of surface waves and nucleoid-associated factors on Escherichia coli. Moreover, light-based modulation has gained significant attention as a potential method for controlling and manipulating proteins. Gill et al.~\cite{gil2020optogenetic} aimed to investigate the modulation of proteins through optogenetics using light-responsive nanobodies that have been developed as a means of regulating proteins in living cells. The system was constructed using a protein of interest, a light-responsive nanobody, and a regulatory component. The binding of the nanobody to the target protein was controlled using a fluorophore that can be switched on and off by a light source. The activity of the target protein was altered using light, resulting in either an increase or decrease in activity. 

Zeng and colleagues examined the impact of light on proteins that naturally form in living cells ~\cite{zeng2022endogenous}. Moreover, Sung et al. conducted research on using optogenetics to activate intracellular antibodies and directly modify endogenous proteins in live cells~\cite{yu2019optogenetic}. They developed a system of intracellular antibodies that can be activated by light. By using this method, intracellular antibodies can more easily attach to the specific targets they are designed to bind to. When these antibodies are exposed to light, they cause a change in the shape of the protein they are targeting. This change can either turn on or turn off the protein's activity.

It has been shown that light may successfully regulate the frequency of proteins in live cells, activating or carrying out specific functions. Furthermore, we can use light of various wavelengths to release or activate small molecules that communicate with proteins, such as enzymes, receptors, ion channels, transcription factors, and kinases.
 
Scientists have been exploring a captivating new area of research in synthetic biology, which involves developing computers using biomolecules~\cite{miyamoto2013synthesizing},~\cite{shapiro2006bringing}.Building and expanding circuits of this nature can be an overwhelming task. DNA-based strand displacement reactions are currently the most powerful molecular circuits that have been discovered. The circuits have certain limitations when it comes to biological systems since it's difficult to genetically encode their components~\cite{schaffter2022cotranscriptionally}. Moreover, researchers tested logic gates on different protein units, including split enzymes and transcriptional machinery, in test tubes, yeast, and primary human T cells. Similar to how logic gates work, each biological process requires a specific combination of proteins to be activated. For instance, in an AND gate, both input proteins need to be present, while in an OR gate, either of the two can activate the downstream process~\cite{singh2020designing}.

The use of light-induced frequency modulation on proteinoids may have a significant impact on photosynthesis. Plants can transform light energy into chemical energy that is used for their metabolic processes and development. Photosynthesis encompasses two primary phases: the light-dependent reactions and the light-independent reactions. The light-dependent reactions include the absorption of light through photosynthetic pigments like chlorophyll. Pigments help in the conversion of light energy to electrons, which are involved in the synthesis of ATP and NADPH. Rubisco enzymes promote the production of organic compounds, including glucose, from carbon dioxide included in the light independent reactions.

There may be some gaps or problems with how light frequency modulation is thought to work with biological molecules like proteinoids~\cite{liu2018modulation}.  The outcome of this method is influenced by  the stability, solubility, toxicity, and bioavailability of biological molecules, as well as the selectivity, affinity, and kinetics of their proteinoid interactions.  Additionally, the depth of light penetration and tissue distribution, as well as the potential for interference with other cellular components, are significant factors to consider.

In the following section we will explore how proteinoids, specifically L-Glu:L-Phe:L-His, L-Glu:L-Phe, L-Phe:L-Lys, L-Phe, and L-Asp, are modulated in terms of frequency.  Additionally, we should review the histograms that display the amplitudes and dominant frequencies. Finally, we will examine the impact of black-white light as well as the effects of solely white or black light on the proteinoids.

\section{Methods}

The amino acids L-Phenylalanine, L-Aspartic acid, L-Histidine, 
L-Glutamic acid and L-Lysine with a purity of more than 98\% were purchased from Sigma Aldrich. The proteinoids were produced using processes described elsewhere~\cite{mougkogiannis2023transfer}. The structure of the proteinoids was observed using scanning electron microscopy (SEM). This was accomplished using the FEI Quanta 650 equipment. Finally, as previously mentioned~\cite{mougkogiannis2023transfer}, Fourier-transform infrared spectroscopy (FT-IR) was utilised to characterise the proteinoids. The light modulation of the proteinoids' spiking frequency was investigated using the light source PHOTONIC PL-2000 (Ryf AG, Switzerland) and UV-light source: QTXLight, Power 230 Vac, 50 Hz, tube length 450 mm. 

Electrical activity of the proteinoids was recorded using pairs of iridium-coated stainless steel sub-dermal needle electrodes (Spes Medica S.r.l., Italy), with twisted cables and  ADC-24 (Pico Technology, UK) high-resolution data logger with a 24-bit A/D converter. Galvanic isolation and software-selectable sample rates all contribute to a superior noise-free resolution. Each pair of electrodes reported a potential difference between the electrodes. In each pair of differential electrodes, the distance between electrodes was c. 10~mm. We recorded electrical activity at one sample per second. During the recording, the logger has been doing as many measurements as possible (typically up to 600 per second) and saving the average value.

\section{Frequency modulation}

\begin{figure}[!tbp]
\centering
\includegraphics[width=0.8\textwidth]{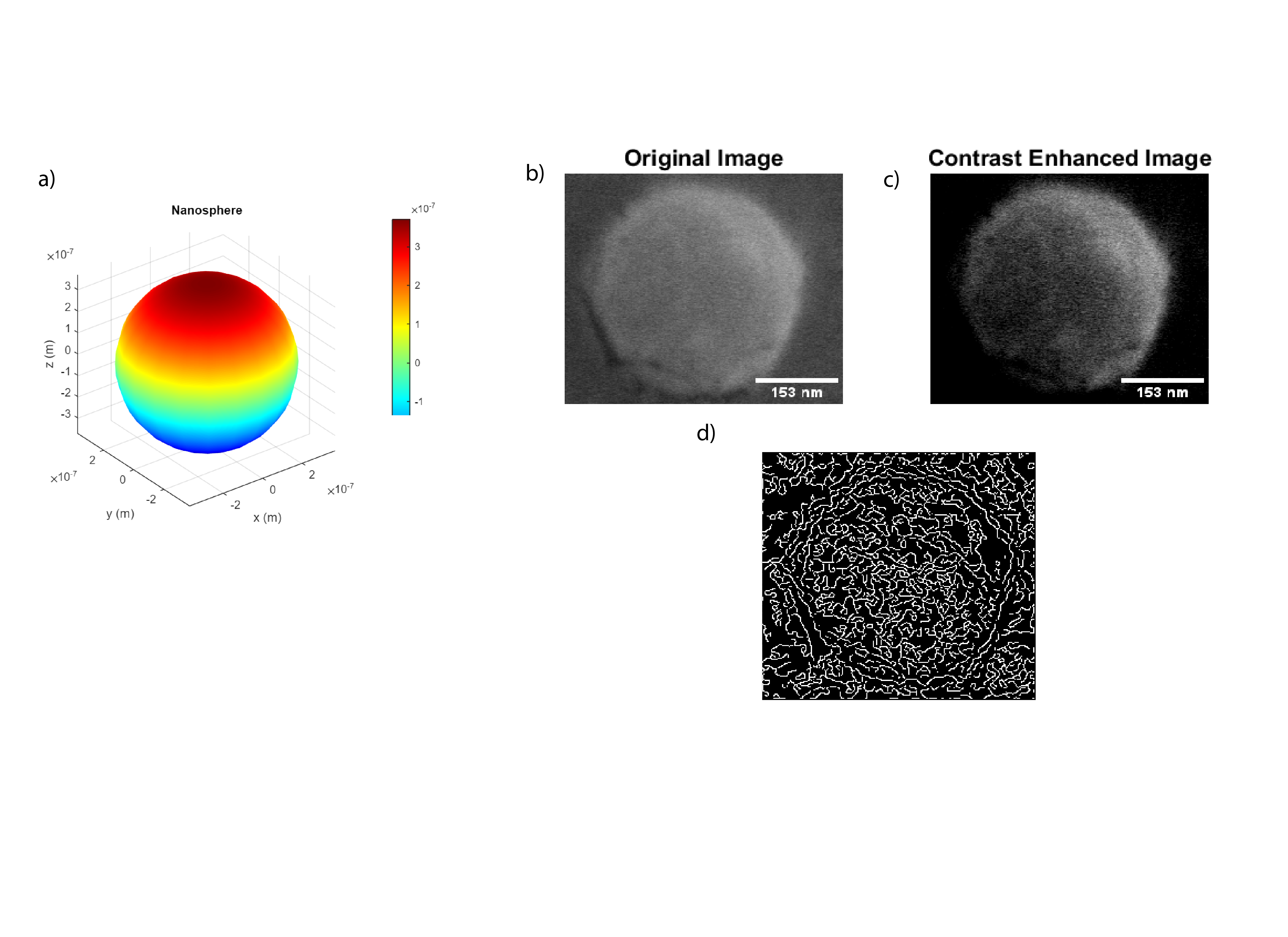}
\caption{The nanospheres of the proteinoid L-Glu:L-Asp:L-Phe have a diameter of 370.4 nm, a surface area of 4.32e-13 $m^{2}$, and a volume of 2.67e-20 $m^{3}$.b) Additionally, a scanning electron microscopy (SEM) image of the nanosphere is provided. c),d) After conducting an analysis on the edge detection of proteinoid nanospheres using Matlab, the results showed a contrast value of 0.1964, a correlation value of 0.8867, an energy value of 0.1928, and a homogeneity value of 0.9018.
}
\label{jn eegbeb bk bkb k}
\end{figure}

Figure~\ref{jn eegbeb bk bkb k} depicts an image captured by a scanning electron microscope (SEM) showcasing a nanosphere of L-Glu:L-Asp:L-Phe in the initial phases of proteinoids replication. The nanosphere is comprised of a spherical arrangement composed of multiple amino acids, including L-glutamic acid, L-aspartic acid, and L-phenylalanine. These amino acids serve as fundamental constituents for the replication of proteinoids. The nanosphere exhibits a diameter of 370.394 nm, a surface area of 4.31$\times$$\mathrm{10^{-13}}$ $\mathrm{m^{2}}$, and a volume of 2.67$\times$$\mathrm{10^{-20}}$ $\mathrm{m^{3}}$. The SEM image of the nanosphere displays a significant level of intricacy, exposing the complex structure of the L-Glu:L-Asp:L-Phe proteinoids that are organised in a spherical arrangement. The process of arranging L-Glu:L-Asp:L-Phe molecules into a three-dimensional configuration within the nanosphere is a pivotal step in the replication of proteinoids. The structure mentioned above is considered to be crucial for the following stage of proteinoid replication, which involves the formation of the proteinoid assembly. 

\begin{figure}[!tbp]
\centering
\includegraphics[width=1\textwidth]{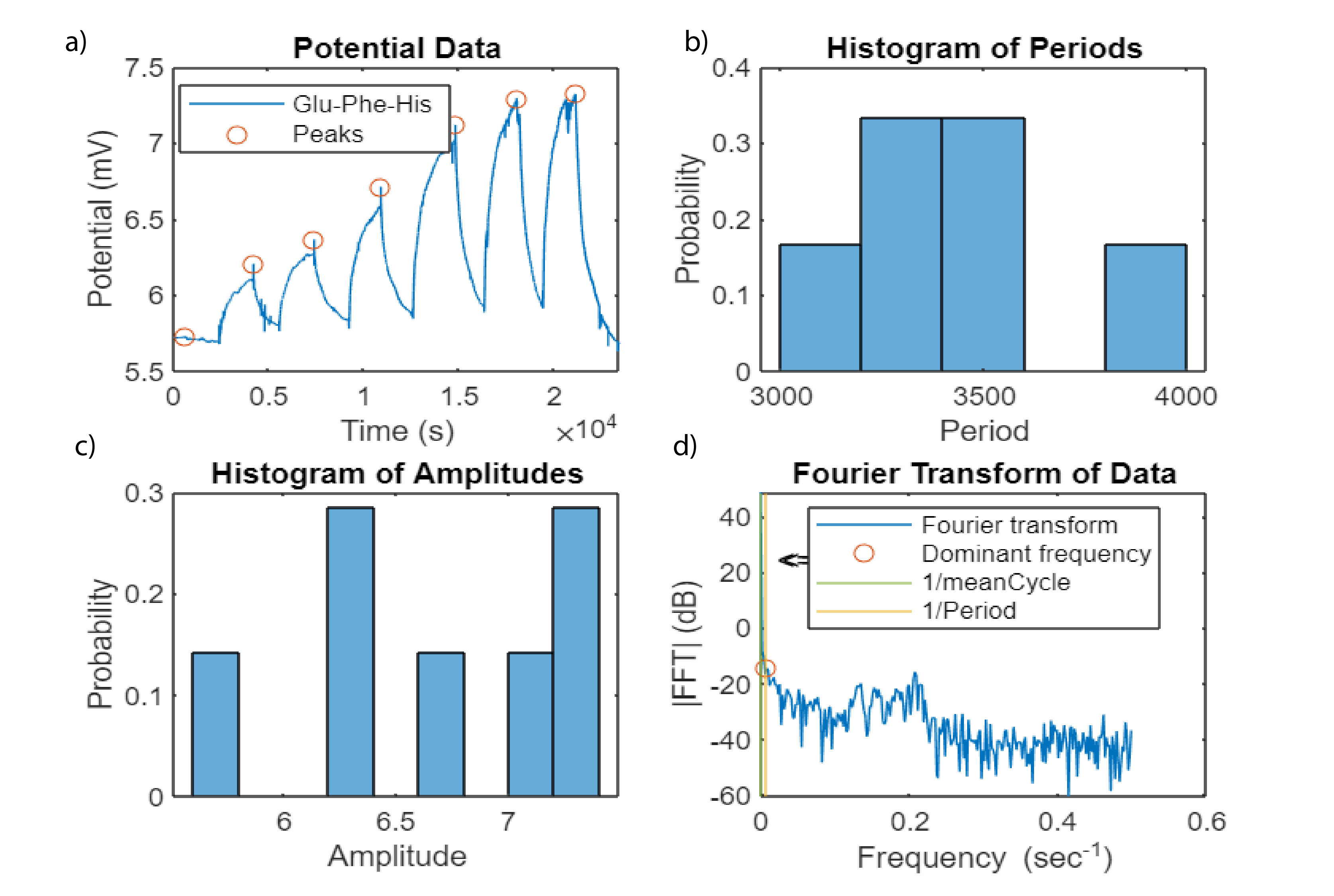}
\caption{The L-Glu:L-Phe:L-His proteinoid oscillations have a period of 128~seconds and an average interval between maxima of 3417.8 seconds, as shown by the a) electrical oscillations, b) histogram of periods, c) histogram of amplitudes, and d) Fourier transform of data. Illumination  with cold white light, from 37.2  to 186.6 klux  (30 min ON - 30 min OFF).
}
\label{ppsoiunpsssazx999}
\end{figure}

\begin{table}[!tbp]
\centering
\caption{Measurements of peak voltage (mV) and periods (sec) were taken before and after illuminating the protenoids L-Glu:L-Phe:L-His, L-Glu:L-Phe,L-Phe:L-Lys, and L-Phe with cold white light at intensities ranging from 37.2 klux to 186.6 klux for 30 minutes on --- 30 minutes off cycles.}
\begin{tabular}{|c|c|c|c|}
	\hline
	Proteinoid  & Light & Peaks & Periods      \\
       &  Condition  &  Mean  &  Mean     \\
    &  & (mV) & (sec)      \\
	\hline\hline
	
	L-Glu:L-Phe:L-His& Periodic & 6.68$\pm$0.61 &  3417.83$\pm$316.04    \\
 L-Glu:L-Phe & Periodic & 2.87$\pm$0.47 & 2174.00$\pm$612.09    \\
 L-Phe:L-Lys & Periodic & 23.77$\pm$0.83 &   3265.83$\pm$451.62   \\
 L-Phe & Periodic & 0.82$\pm$0.43 & 759.32$\pm$108.34      \\
	\hline
	\end{tabular}
\label{vdbfngdm,ukt.il.o/o/o;}
\end{table}

\begin{table}[!tbp]
\centering
\caption{The table displays the mean ($\mu$), standard deviation ($\sigma$), and negative log likelihood (NLL) of functions fitted to the standard distribution for L-Glu:L-Phe:L-His,L-Glu:L-Phe,L-Phe:L-Lys,L-Asp,L-Phe proteinoids.
}
\begin{tabular}{|c|c|c|c|}
	\hline
	Proteinoid  & $\mu$ & $\sigma$ &       \\
       &  Mean  &  Std  & NLL      \\
    &  &  &     \\
	\hline\hline
	
	L-Glu:L-Phe:L-His& 3247.9 & 760.83  & 148.06   \\
 L-Glu:L-Phe & 3534.3 & 453.94 &  272.21   \\
 L-Phe:L-Lys & 3742.9  & 517.55 &   248.54    \\
 L-Phe & 3400.8 & 1144.8   & 122.58      \\
 L-Asp & 2237.4 & 745.87  & 118.03      \\
	\hline
	\end{tabular}
\label{dnckscks}
\end{table}

Proteinoids respond to illumination with spikes of electrical potential as illustrated in Fig.~\ref{ppsoiunpsssazx999}a. Theses spikes of electrical potential are characterised in Tab.~\ref{vdbfngdm,ukt.il.o/o/o;}. Tables~\ref{vdbfngdm,ukt.il.o/o/o;} and ~\ref{dnckscks} present a quantitative summary of the average pk and average periods of the proteinoids L-Glu:L-Phe:L-His, L-Glu:L-Phe, L-Phe:L-Lys, L-Phe, and L-Asp. The average peak values exhibit a range of $-$13.55$\pm$5.06 to 23.77$\pm$0.83~mV, whereas the average periods span from 759.32$\pm$108.34 to 6272.88$\pm$16.84~sec. 

\begin{figure}[!tbp]
\centering
\includegraphics[width=1\textwidth]{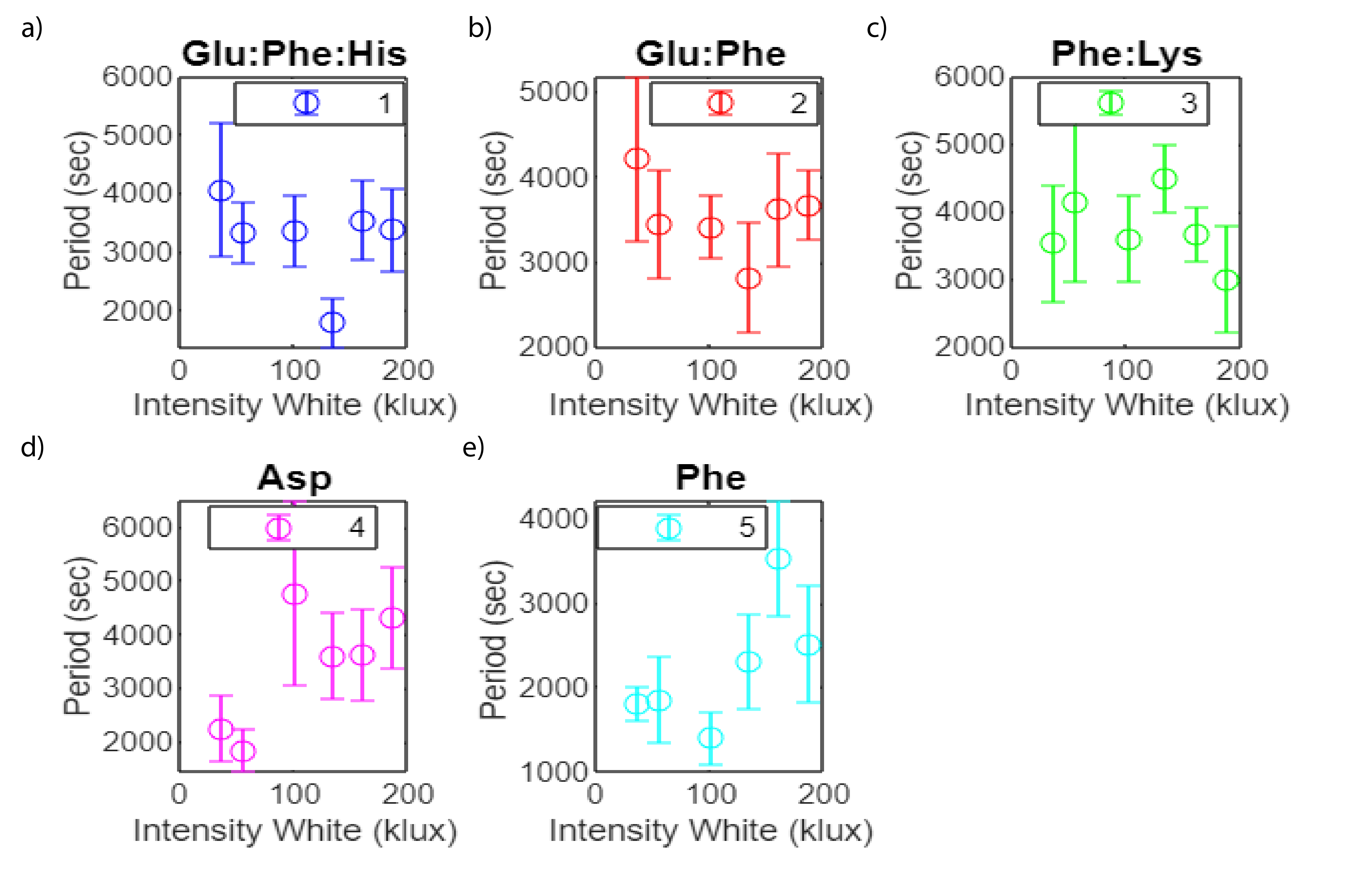}
\caption{Diagram displaying the period in seconds of proteinoids with codes 1 (L-Glu:L-Phe:L-His),2 (L-Glu:L-Phe),3 (L-Phe:L-Lys),4 (L-Asp), and 5 (L-Phe) in response to varying white light intensities in klux.
}
\label{kfmdklngkbsbb}
\end{figure}

\begin{figure}[!tbp]
\centering
\includegraphics[width=1\textwidth]{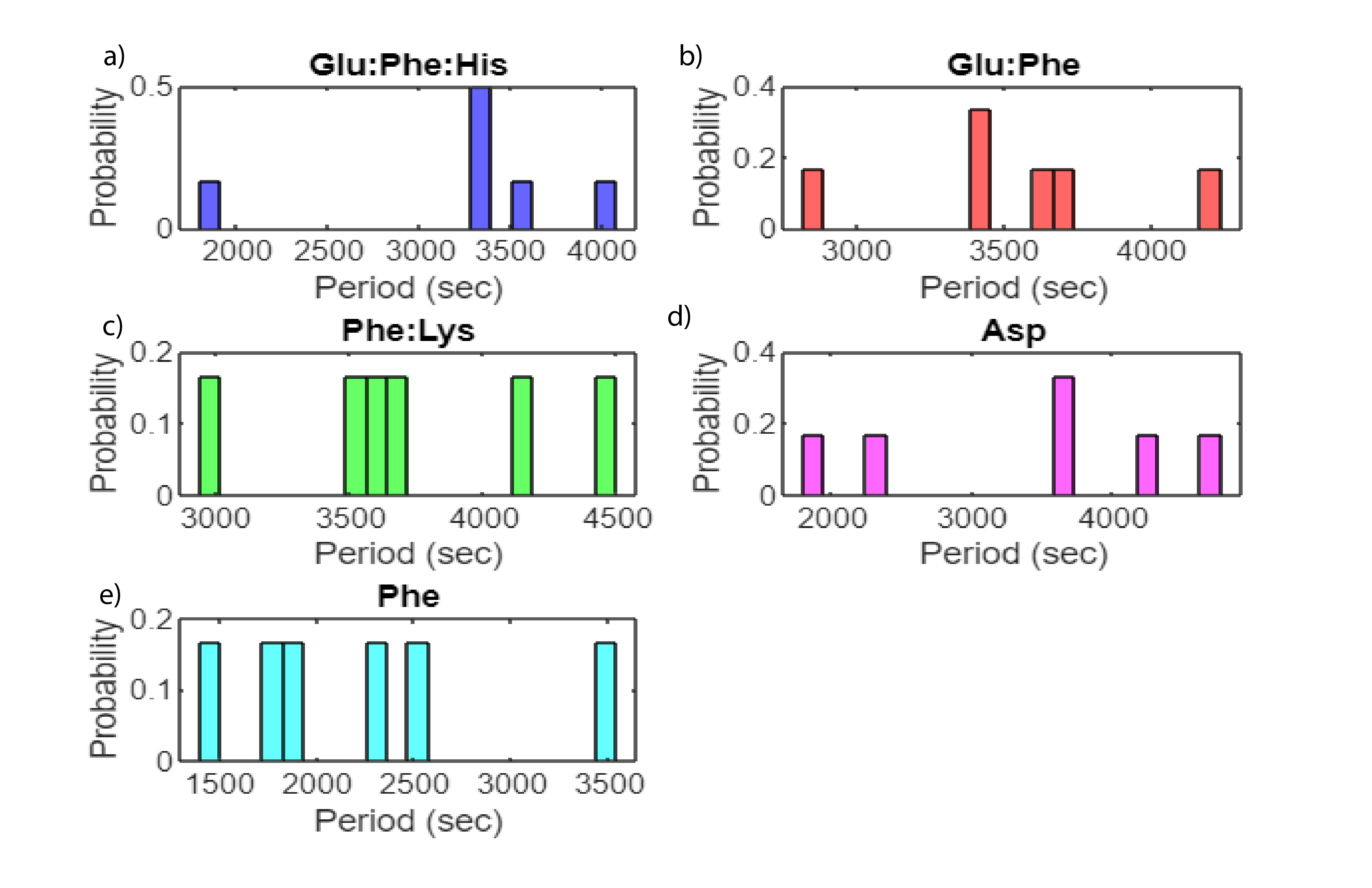}
\caption{Diagram displaying the histogram of period in seconds of proteinoids with codes 1 (L-Glu:L-Phe:L-His),2 (L-Glu:L-Phe),3 (L-Phe:L-Lys),4 (L-Asp), and 5 (L-Phe) in response to varying white light intensities in klux.
}
\label{asdvsabvdbfdbd}
\end{figure}

The outcomes demonstrate that there is no significant difference in the mean peak and mean periods among the various proteinoids. Subsequently, the histograms depicting the amplitudes serve to emphasise the dominant frequency, peaks, and spectrum of frequencies pertaining to the proteinoids (Figs.~\ref{kfmdklngkbsbb} and \ref{asdvsabvdbfdbd}). The histograms illustrate that proteinoids exhibit a prevailing period of 80 to 180 seconds, with a higher percentage of periods in the 120-second interval. The present study aims to analyse the impact of black-white light and monochromatic light (white or black) on proteinoids. After being subjected to black-white light for a duration of 30 minutes, the proteinoids exhibited a rise in their average pk values, with the highest point being recorded at 8.33 kHz. Under the condition of exposure solely to white light, the average peak values exhibited an increase while the periods remained constant. Under the exclusive exposure to black light, the average pk values exhibited a decline, reaching a maximum at 3.53~kHz, while the average periods remained consistent.

When proteins are exposed to alternating periods of 30 minutes of black light, followed by 30 minutes of white light, their periodicity can be analysed using a distribution function that has been fitted to the data.

The tripeptide L-Glu:L-Phe:L-His exhibits a mean value of 3247.9~sec, a standard deviation of 760.8, and a negative log likelihood (NLL) of 148.1. The mean, standard deviation, and negative log likelihood (NLL) of proteinoid L-Glu:L-Phe are 3534.3 sec, 453.9, and 272.2, respectively. L-Phe:L-Lys displays a mean of 3742.9 sec, a standard deviation of 517.5, and an NLL of 248.5. L-Phe exhibits a mean of 3400.8 sec, a standard deviation of 1144.8, and an NLL of 122.6. Lastly, L-Asp has a mean of 2237.4 sec, a standard deviation of 755.9, and an NLL of 118. The proteinoids' normal distribution implies that each compound possesses a unique range of variability in its properties. Specifically, proteinoid L-Asp displays the narrowest range, while proteinoid L-Phe exhibits the widest range. The observed variations in the mean, standard deviation, and negative log-likelihood (NLL) values of the proteinoids imply that each proteinoid exhibits unique properties, which may be indicative of varying structural conformations or degrees of stability.

\begin{figure}[!tbp]
\centering
\includegraphics[width=1\textwidth]{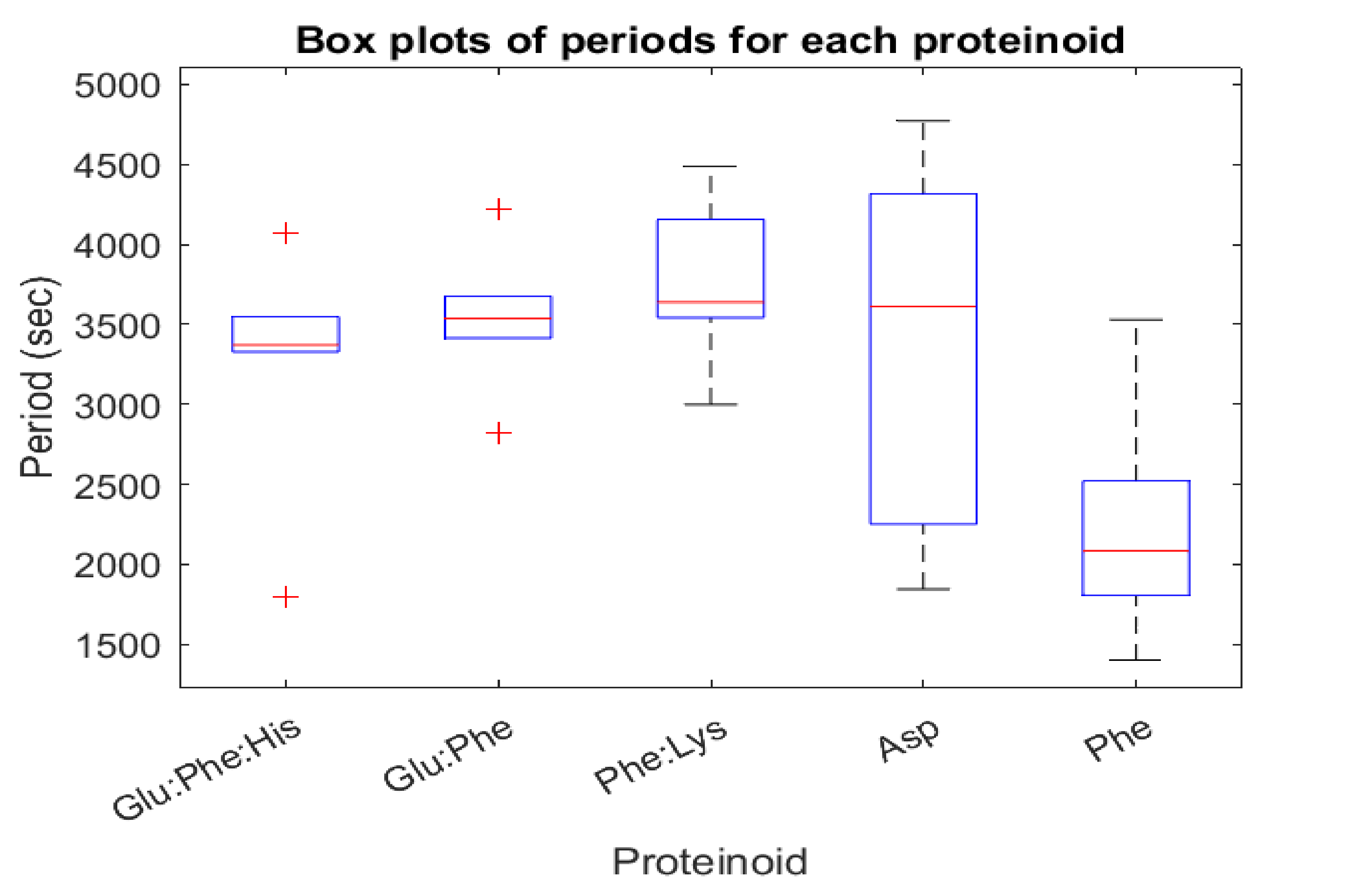}
\caption{The box plot demonstrates the distribution of periods for L-Glu:L-Phe:L:His, L-Glu:L-Phe, L-Phe:L-Lys, L-Asp, and L-Phe. The 25th and 75th percentiles for L-Glu:L-Phe:L-His are respectively 3328.64 and 3548 seconds. The 25th and 75th percentiles for L-Glu:L-Phe are, respectively, 3412.75 and 3676.57 seconds. The 25th and 75th percentiles for L-Phe:L-Lys are, respectively, 3541.38 and 4154.73.57 seconds. The 25th and 75th percentiles for L-Asp are, respectively, 2251.13 and 4316.33 seconds.
 The 25th and 75th percentile times for L-Phe are, respectively, 1807.8 and 2529.9 seconds. Periodic simulations of proteinoids were conducted using cold white light (37.2-186.6 klux) and black light (695.8 lux) for a duration of 30 minutes during the measurement process.}
\label{sdgfahdsgndhmj,,}
\end{figure}

The boxplot results depicted in Fig.~\ref{sdgfahdsgndhmj,,} indicate that the periods median of the proteinoids L-Glu:L-Phe:L-His, L-Glu:L-Phe, L-Phe:L-Lys and L-Asp are uniformly distributed at 3400 seconds, whereas L-Phe exhibits a median of approximately 2200 seconds. The results depicted in Fig.~\ref{sdgfahdsgndhmj,,} indicate that the proteinoids L-Glu:L-PheL-His, L-Glu:L-Phe, L-Phe:L-Lys, and L-Asp exhibit a significantly high degree of hydrophilicity, as evidenced by their respective period times clustering around the 3400~sec mark. In contrast, L-Phe exhibits lower hydrophilicity, as evidenced by its period time of approximately 2200~sec. The aforementioned data possesses the potential to facilitate comparisons and differentiation among proteinoids, given that the retention time of a proteinoid is subject to the hydrophobicity exhibited by its constituent amino acids. In the event that a proteinoid possessed an amino acid composition identical to the four proteinoids depicted in Fig.~\ref{sdgfahdsgndhmj,,}, but with L-Asp replacing L-Phe, it is expected that the mean period would exhibit a significant decrease. This observation suggests that the hydrophilicity of the proteinoid in question is comparatively higher than that of the remaining four proteinoids, thus potentially enabling its differentiation from the others.

The Fast Fourier Transform (FFT) was utilised to decompose the signal into its frequency components. The signal can be decomposed into its frequency components by plotting the magnitude of the FFT output against frequency, as depicted in Fig.~\ref{ppsoiunpsssazx999}. This form of analysis proves to be beneficial in comprehending the behaviour of the signal in the frequency domain, specifically the quantification of energy present in each frequency component. The analysis can also yield valuable information regarding the frequency composition of the signal, including the presence of specific frequencies and their relationships.

Typically, the firing rate is determined by calculating a temporal average, as illustrated in Figure~\ref{sdnanbvbvkjhgh,hja;b}. To conduct the experiment,we select a specific time frame, such as T = 100 ms or T = 500 ms. During this time frame, we will record the number of spikes that occur, which is denoted as $n_{sp}(T)$. When we divide the number of times a neuron fires by the length of the time window, we get the mean firing rate (Eq.~\ref{vsnkdngjsadn}). When discussing firing rate, it is common to express it in Hz units.
For the past century, the idea of mean firing rates has been effectively utilised. When a muscle is stretched, the neurons in the muscle's stretch receptors fire at a rate that is directly related to the amount of force being applied to the muscle. Over the next few decades, scientists began to use firing rates as a common method for characterising the behaviour of various types of cortical or sensory neurons~\cite{gerstner2002spiking}.

\begin{equation} \label{vsnkdngjsadn}
\nu = \frac{n_{sp}(T)}{T}
\end{equation}

\section{Realisation of Boolean gates}

\begin{table}[!tbp]
    \centering
    \begin{tabular}{c|cc}
    spikes    & gate  & notations   \\  \hline
\includegraphics[scale=0.2]{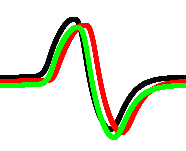}  & {\sc or} & $x+y$ \\
\includegraphics[scale=0.2]{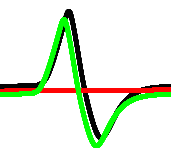}  & {\sc select} & $y$ \\
\includegraphics[scale=0.2]{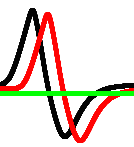}  & {\sc xor} & $x \oplus y$ \\
\includegraphics[scale=0.2]{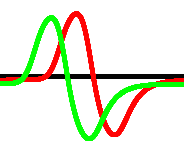}  & {\sc select} & $x$ \\
\includegraphics[scale=0.2]{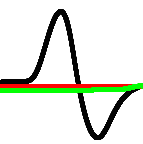}  & {\sc not-and} & $\overline{x}y$ \\
\includegraphics[scale=0.2]{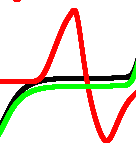}  & {\sc and-not} & $x\overline{y}$ \\
\includegraphics[scale=0.2]{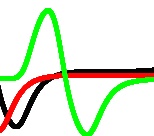}  & {\sc and } & $xy$ \\
    \end{tabular}
    \caption{Representation of gates by combinations of spikes. Black lines show the potential when the network was stimulated by input pair (01), red by (10) and green by (11).}
    \label{tab:spikes2gates}
\end{table}

We have detected Boolean gates using the technique proposed in \cite{adamatzky2019computing}. We here assume that each spike represents logical {\sc True} and that spikes occurring within less than 3000~sec of each other happen simultaneously. Then a representation of gates by spikes and their combinations will be as shown in Tab.~\ref{tab:spikes2gates}. 

\begin{figure}[!tbp]
\centering
\includegraphics[width=1\textwidth]{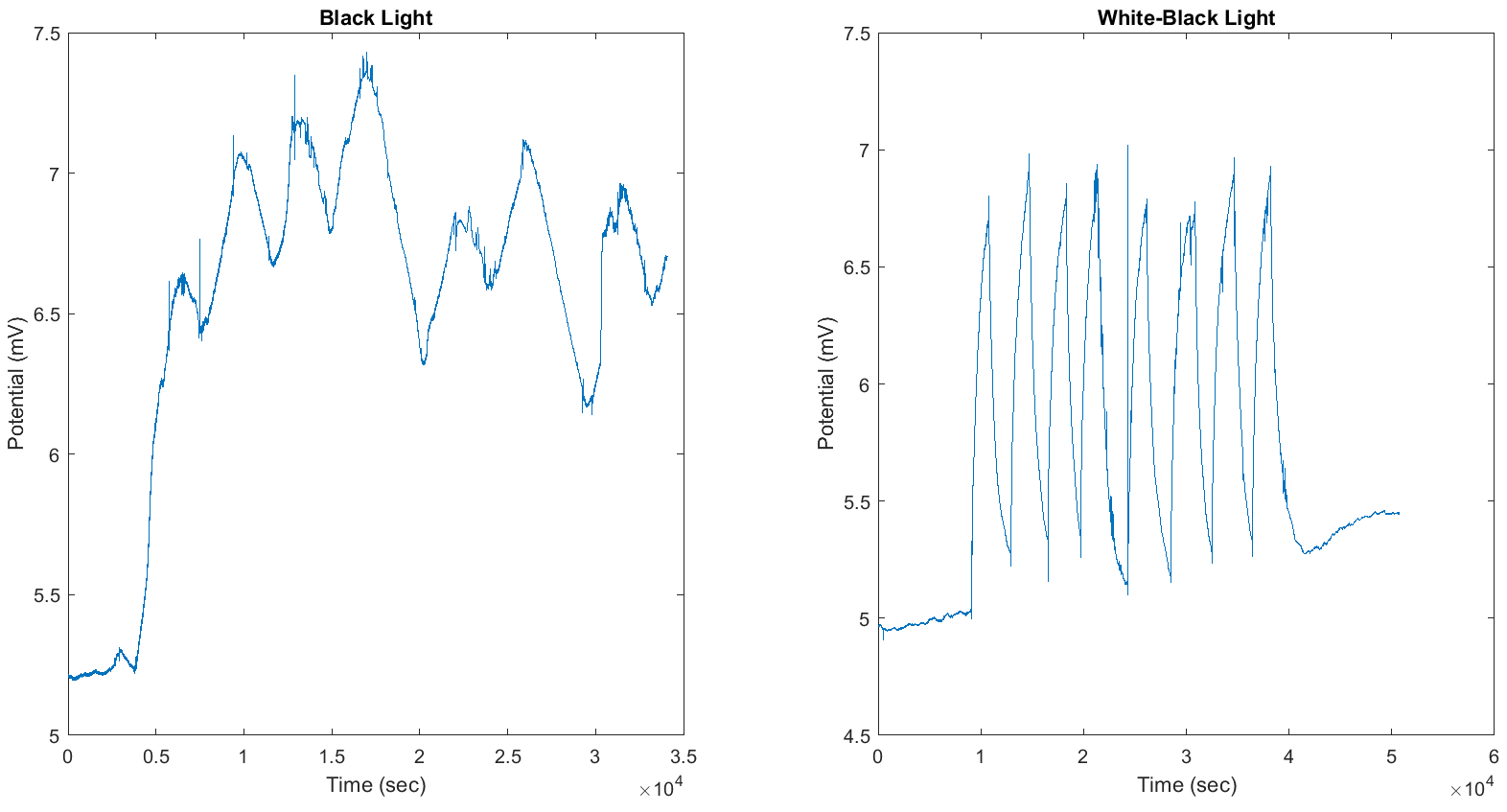}
\caption{Left: Spikes when the proteinoid exposed to a cycle of black light and no light for every 30 min. Right: spiking of proteinoid L-Glu:L-Phe:L-His when exposed to a cycle of white and black light for 30 min. }
\label{sdnanba;b}
\end{figure}

\begin{figure}[!tbp]
\centering
\includegraphics[width=1\textwidth]{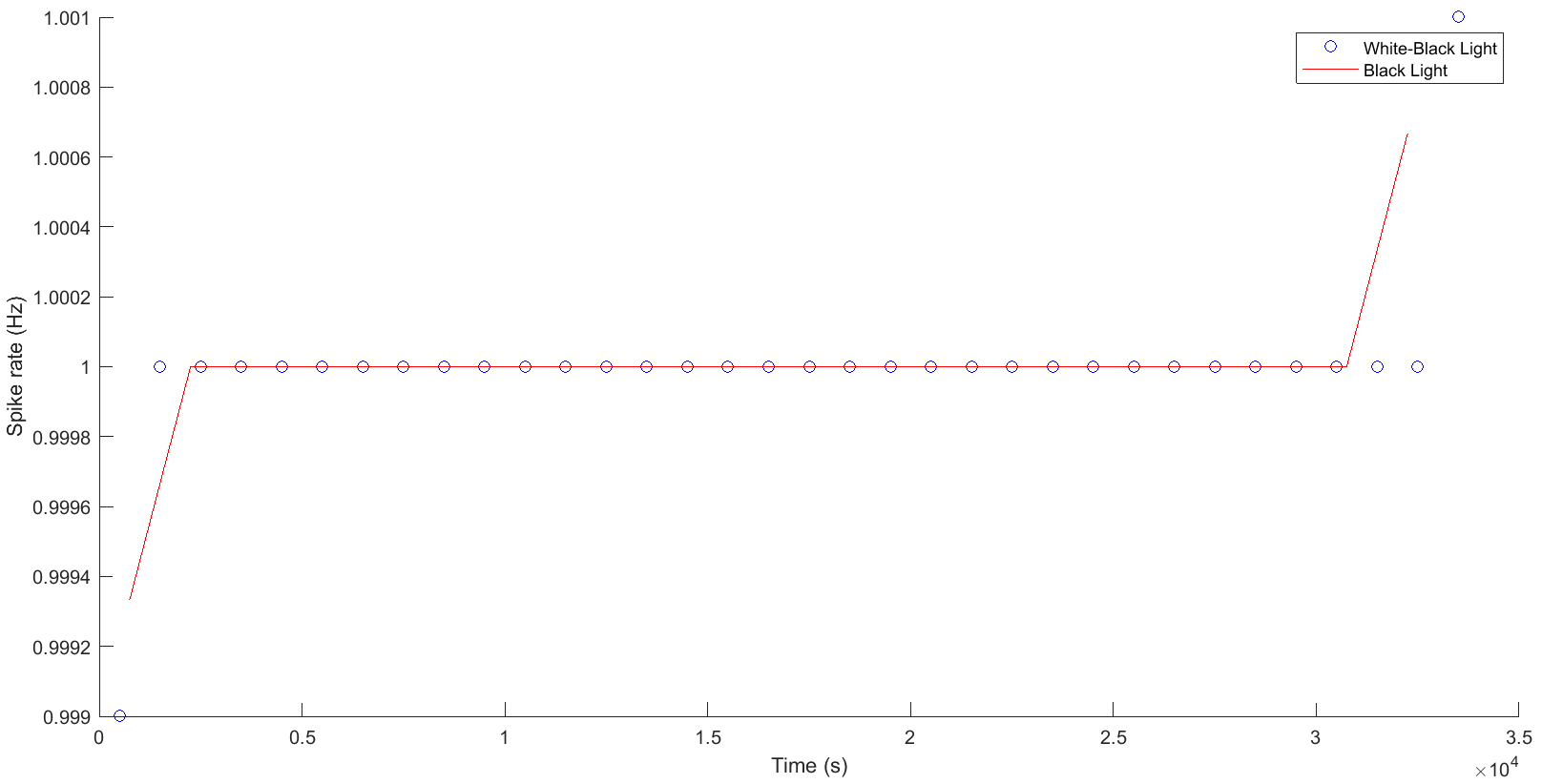}
\caption{ The graph displays the projected spike rates for the proteinoid L-Glu:L-Phe:L-His over time. The spike rate for the proteinoid under white-black light illumination is represented by the blue points, while the spike rate for the proteinoid under black illumination is represented by the red line. To determine the spike rate, we count the number of spikes that occur within a given time interval, or "bin," and then divide that count by the duration of the bin. The size of the bin is a factor that impacts both the accuracy and the level of interference in the estimation of spike rate. When using a smaller bin size, you can achieve higher resolution, but it comes at the cost of increased noise. When using a larger bin size, the resolution of the data decreases, but it also results in less noise. According to the graph, it appears that the spike rates for both illuminations are quite similar, although not exactly the same. The spike rates exhibit some fluctuations, with peaks and valleys that could potentially be attributed to the stimulus or other variables~\cite{cajigas2012nstat,NSChapter,Preprocessing}. }
\label{scsfsdvsdvsvsasaccvazvv;b}
\end{figure}

\begin{figure}[!tbp]
\centering
\includegraphics[width=1\textwidth]{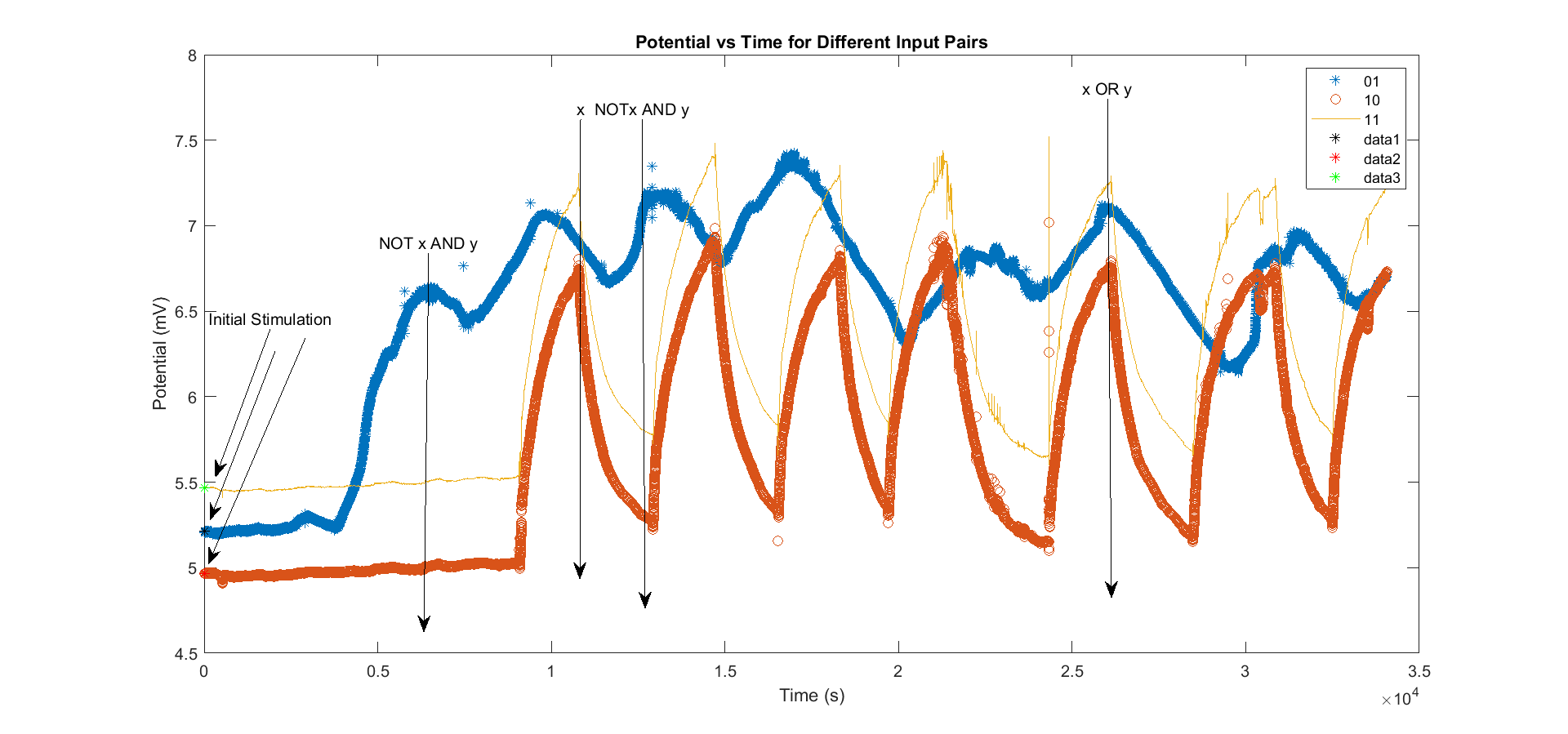}
\caption{This illustration showcases the use of spikes for the purpose of implementing logic gates. For each input pair (01, 10, and 11), there are potential values assigned to them which are referred to as data1, data2, and data3. The data points are graphed on a single set of axes, using various colours and line styles to differentiate between them. Each input pair is marked by a star to indicate the moment of initial stimulation.}
\label{sdnanbvbvkjhgh,hja;b}
\end{figure}

\begin{figure}[!tbp]
\centering
\includegraphics[width=0.8\textwidth]{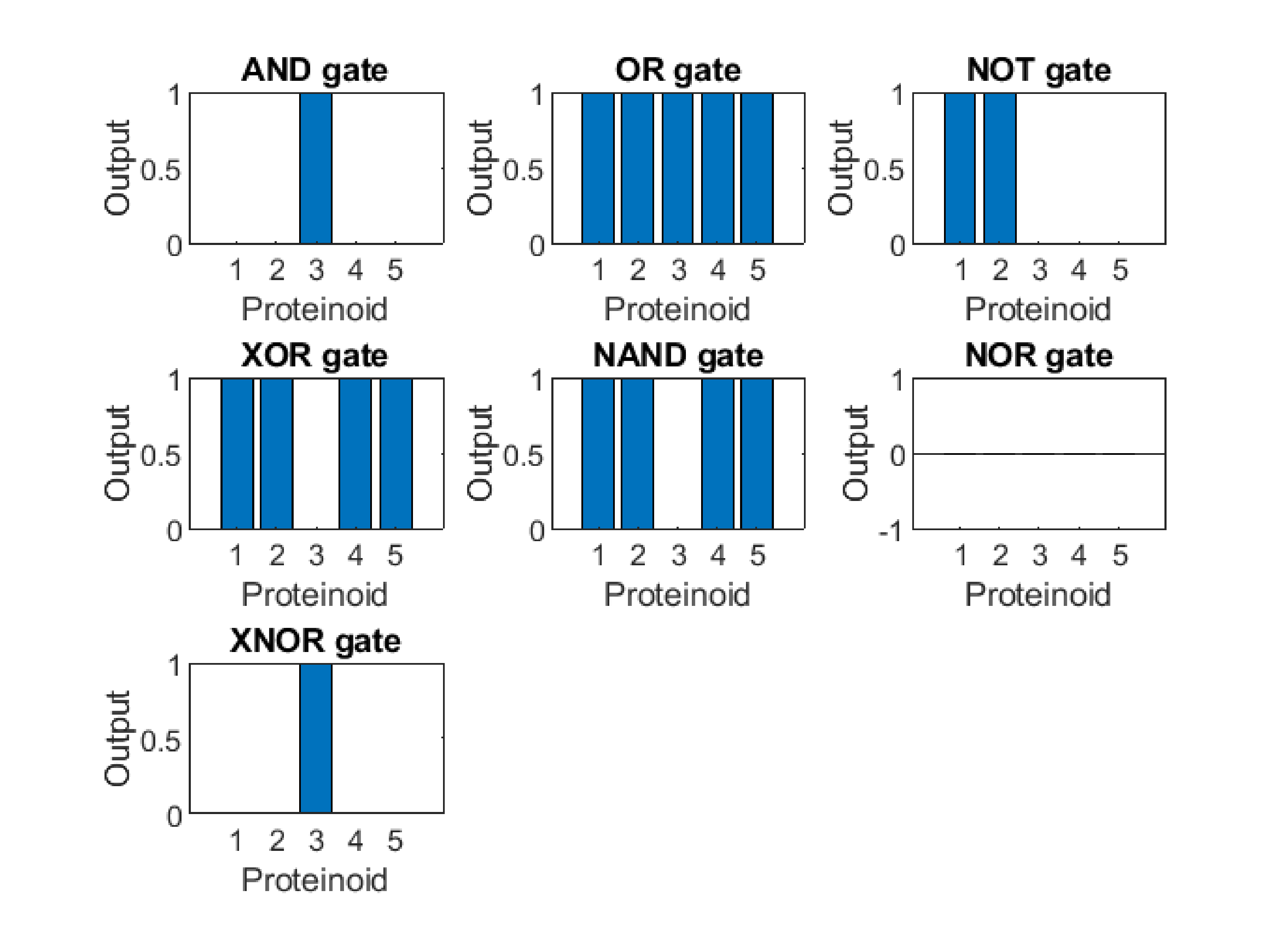}
\caption{This figure examines the values of various logical operators, including AND, OR, NOT, XOR, NAND, NOR, and XNOR. The results are based on the periodic illumination of proteinoids with black and black-white light (two inputs: white, black-white light), and the magnitude of the periods is measured in seconds. The plot specifies two vectors of periods in seconds for black light and black and white light. It defines a threshold for spikes and periods without spikes, a value that divides the periods into two categories. Based on the threshold, it allocates the value 0 to no spikes and 1 to spikes. It uses the vectors of spikes as input for the preceding code, which creates and plots the AND, OR, NOT, XOR, NAND, NOR and XNOR logic gates for the inputs and outputs. Here, Proteinoid 1: L-Glu:L-Phe:L-His, 2: L-Glu:L-Phe, 3: L-Phe:Lys, 4: L-Phe,5: L-Asp. }
\label{afsdgabdsnsngsnsj}
\end{figure}

\begin{figure}[!tbp]
\centering
\includegraphics[width=0.8\textwidth]{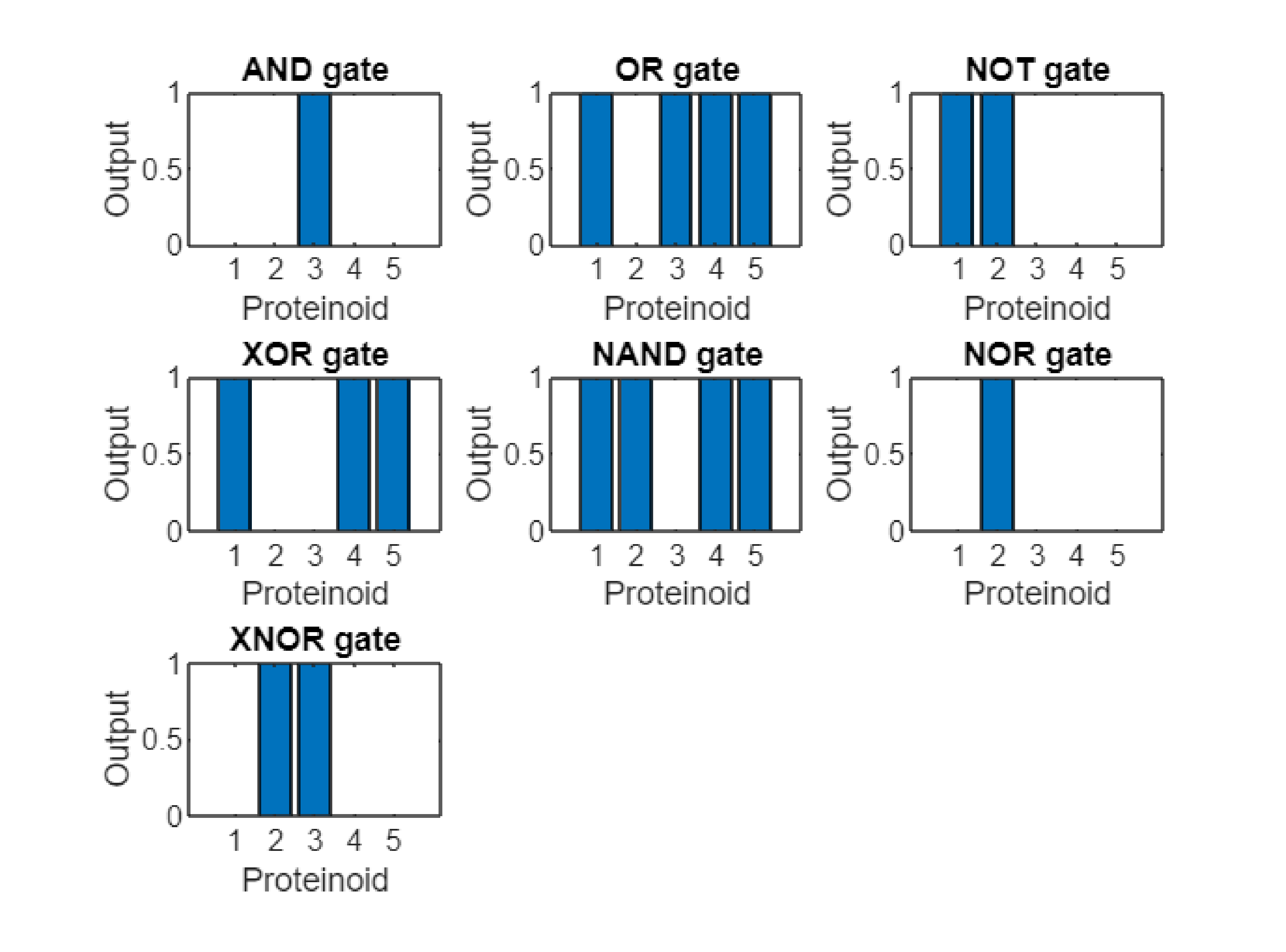}
\caption{ This figure examines the values of various logical operators with two inputs (white, black light), including AND, OR, NOT, XOR, NAND, NOR, and XNOR. The results are based on the periodic illumination of proteinoids with black and white light, and the magnitude of the periods is measured in seconds. It uses the vectors of spikes as input for the preceding code, which creates and plots the AND, OR, NOT, XOR, NAND, NOR and XNOR logic gates for the inputs and outputs. Here, Proteinoid 1: L-Glu:L-Phe:L-His, 2: L-Glu:L-Phe, 3: L-Phe:Lys, 4: L-Phe,5: L-Asp. }
\label{gkjhvgghvhcgc}
\end{figure}

\begin{figure}[!tbp]
\centering
\includegraphics[width=0.8\textwidth]{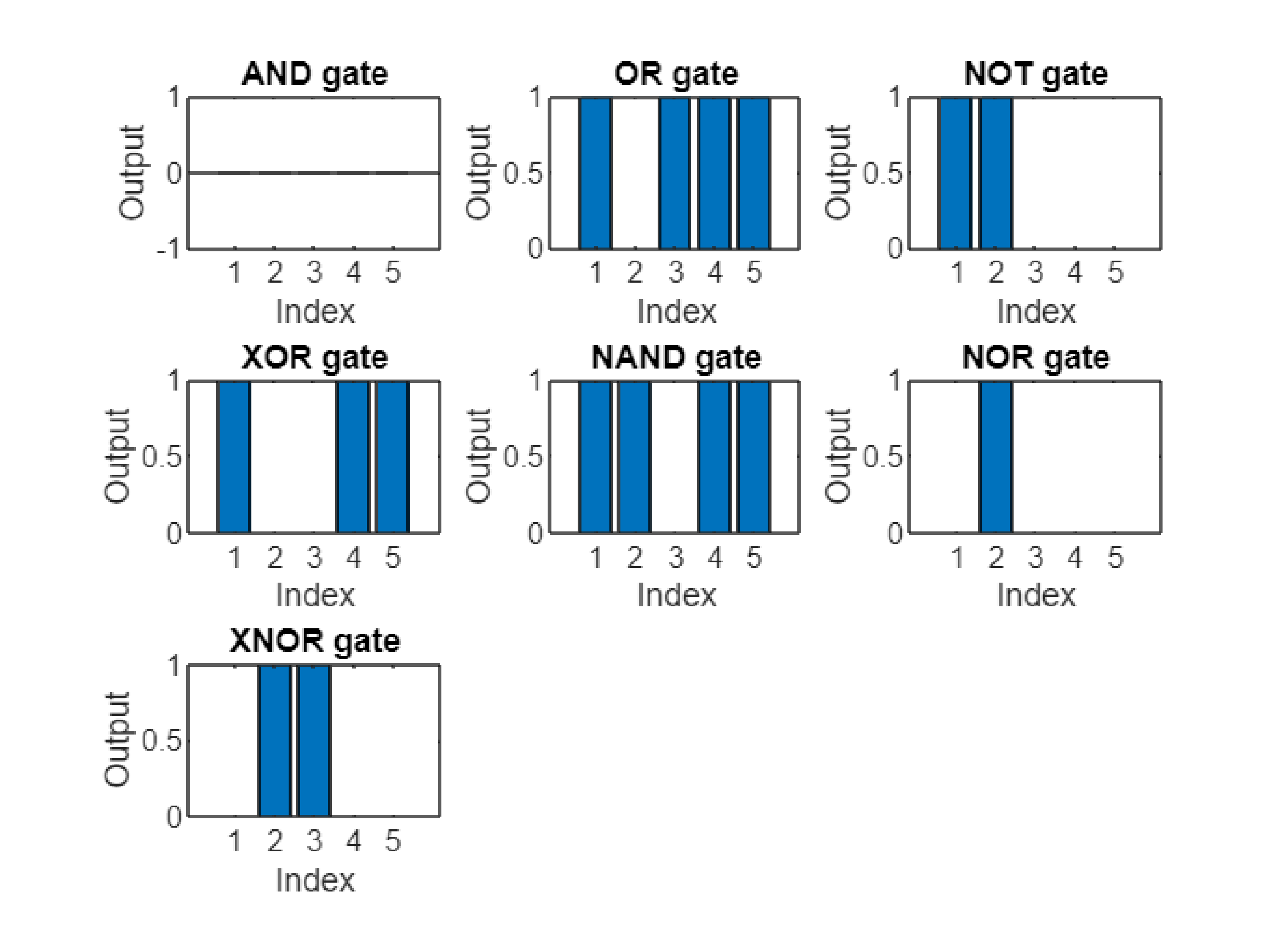}
\caption{ This figure examines the values of various logical operators with three inputs (white, black, no light), including AND, OR, NOT, XOR, NAND, NOR, and XNOR. The results are based on the periodic illumination of proteinoids with black and white light, and the magnitude of the periods is measured in seconds.It uses the vectors of spikes as input for the preceding code, which creates and plots the AND, OR, NOT, XOR, NAND, NOR and XNOR logic gates for the inputs and outputs. Here, Proteinoid 1: L-Glu:L-Phe:L-His, 2: L-Glu:L-Phe, 3: L-Phe:Lys, 4: L-Phe,5: L-Asp. }
\label{dsgagdahadn}
\end{figure}

Figure~\ref{afsdgabdsnsngsnsj} depicts the conversion of periods into logical gates through thresholding of proteinoids periods, subsequent to being exposed to periodic black and white light. The threshold value is determined to be 3423 seconds. The present study involves the execution of logical operations on logical values, including logical AND, logical OR, logical NOT, logical XOR, logical NAND, logical NOR, and logical XNOR. Several conclusions can be inferred from Figure~\ref{afsdgabdsnsngsnsj}, including the following: The selection of the threshold value ought to be based upon certain criteria or statistics pertaining to the periods, such as the mean, median, or standard deviation.
In order to prevent all logical values from being 0, it is advisable for the threshold value to be set lower than the maximum value of the periods.
When all logical values are 0, the logical operations lack 
significance.

The plot ~\ref{afsdgabdsnsngsnsj} illustrates how the input and output of various logic gates are related. A logic gate is a fundamental electronic component that carries out a simple logical operation on one or more binary inputs and generates a single binary output. An AND gate is a logic gate that produces an output of 1 only when both of its inputs are 1. If either or both inputs are 0, the output will be 0. When you use a NOT gate, the output it produces is the exact opposite of the input it receives. This means that if the input is 0, the output will be 1, and if the input is 1, the output will be 0. Finally, logic gates are designed with unique truth tables that display the output for every possible input combination.

Figure~\ref{sdnanba;b} displays a comparison of the potential versus time for two distinct light sources --- white-black light and black light. The amount of potential is typically quantified in mV, while the duration of time is typically measured in seconds. In the diagram labelled as Fig.~\ref{sdnanbvbvkjhgh,hja;b}, you can see how gates are represented through the combination of spikes. For each input pair (01, 10, and 11), there are potential values assigned to them which are data1, data2, and data3. The data is displayed on a single graph with various colours and line styles, all plotted against the time vector. Each input pair is marked by a star, indicating the exact moment of initial stimulation.In the plot, you can see the presence and absence of spikes for each light source. This is determined by a threshold value of 3000~sec. When the time exceeds the threshold, a spike is given a value of 1. On the other hand, if the time is less than or equal to the threshold, a no spike is assigned a value of 0. Logic gates rely on two types of inputs: spikes and no spikes. These inputs are binary values that are used to perform Boolean operations. Essentially, logic gates are functions that take these inputs and produce a corresponding output based on the specific operation being performed. The plot displays how each logic gate responds based on the inputs it receives. The AND gate will only produce an output of 1 if both of its inputs are ``spikes". On the other hand, the OR gate will produce an output of 1 if either one of its inputs is a ``spike". The NOT gate, as the name suggests, will produce the opposite of its input 1. These are just a few examples of how these gates function.

\section{Discussion}

The capacity of mammals to comprehend diverse frequencies characterised by varying amplitudes and periods has garnered the interest of researchers~\cite{charlton2017function}. The process of frequency modulation is responsible for the recognition of various sounds~\cite{Frequency}. The phenomenon of frequency modulation pertains to the capacity of mammals to modify the frequency of a signal for the purpose of enhancing their comprehension of the conveyed information~\cite{wang2022chemically}. Bats utilise echolocation to interpret distinct signals and locate their prey. Researchers have endeavoured to replicate the signal comprehension and differentiation capabilities of mammals through the utilisation of bio-hybrid neuro-interfaces.  Chemical means can be employed by these devices to establish communication with biological neurons.  In addition, there have been advancements in the development of silicon-based artificial neurons that aim to emulate the functionality of authentic neurons found in the respiratory system and hippocampus~\cite{abu2019optimal}. 

The modulation of light can be classified into two distinct categories: frequency modulation and amplitude modulation. The first scenario involves utilising a light beam to convey information through alterations in its frequency, while the following one involves modifying the signal's intensity and amplitude for the same purpose. Frequency modulation (FM) can be employed in various methodologies, including spectroscopy. In this technique, the absorption of light by a material is measured as a function of the modulated frequency, which is altered by changing the refractive index of the modulated material. Frequency modulation (FM) can be accomplished through the utilisation of acousto-optic modulators. The modulation of the refractive index of a material is achieved through the utilisation of acoustic waves.  Frequency Modulation (FM) finds applications in the fields of optical communications and unconventional computing~\cite{oberson1999frequency},~\cite{hu2021chip},~\cite{ziegler2020novel}~\cite{Liu2016Optical}.

The utilisation of black and white cold light enabled the regulation of proteinoid frequency and manipulation of their characteristics. The initial stage of our attempt involved the exploration of novel methods for computation through the utilisation of the characteristics inherent in living organisms. The manipulation of proteinoids through alternating exposure to white and black light is a promising avenue for the development of logic gates with the ability to execute complex computational operations at a high level of performance. 

The use of frequency modulation (FM) is a method employed to regulate the luminance levels within proteinoids. The methodology uses diverse light wavelengths to modulate the amount of energy that is assimilated or discharged by the proteinoids. The underlying principle of frequency modulation is relatively straightforward. As the frequency of electromagnetic radiation rises, there is a corresponding increase in the amount of energy absorbed by the proteinoid that it traverses. On the other hand, a decrease in luminous flux leads to a proportional reduction in the quantity of energy assimilated.

A combination of various wavelengths of light is used to modulate its frequency. The methodology employs a blend of filters to produce diverse wavelengths of electromagnetic radiation. By adjusting the frequency of the combined light, one can modulate energy absorption or emission. Various wavelengths of light are utilised to modulate luminosity. The frequencies of light vary depending on their respective colours. Blue light possesses a higher frequency in comparison to red light, resulting in an enhanced absorption of energy. It can be observed that the frequency of yellow light is comparatively lower than that of green light, resulting in a lower amount of energy being absorbed.

\begin{figure}[!tbp]
\centering
\includegraphics[width=1\textwidth]{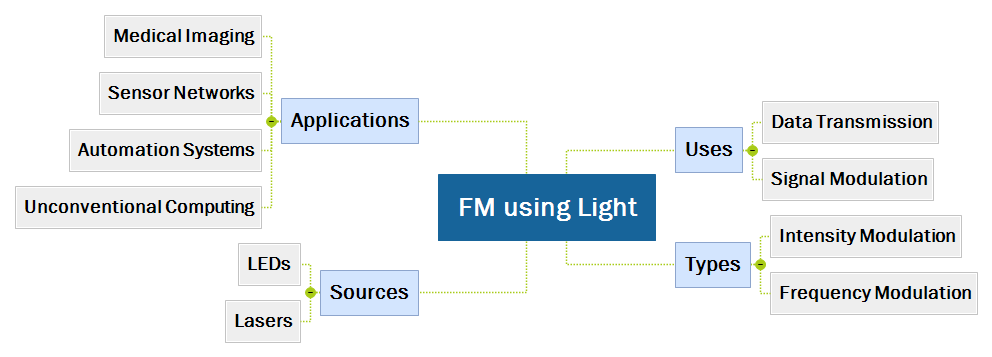}
\caption{The mindmap illustrates applications, sources, types, and uses of frequency modulation of proteinoids.
}
\label{vsgfsdbdn fsn}
\end{figure}

Light-induced frequency modulation of proteinoids has the potential to facilitate the creation of innovative biosensors characterised by heightened specificity and sensitivity. This technology exhibits potential for detecting diverse proteins in the bloodstream or diagnosing a range of medical conditions. Subsequently, the utilisation of light-induced frequency modulation of proteinoids holds potential for targeted drug delivery to precise tissues or cells within the human body. Thirdly, significant progress could be made in the field of tissue engineering. Examples of biological processes that involve the development and activation of novel vascular structures. In addition, the utilisation of light-based frequency modulation has the potential to facilitate the restoration of impaired tissues and promote the proliferation of growing dermal cells. Proteinoid FM has the potential to serve as a means of cancer cell tracking and the development of novel cancer treatment modalities. The present mindmap (Fig.~\ref{vsgfsdbdn fsn}) provides a summary of the key characteristics of proteinoids that can be utilised through the application of frequency modulation.

\begin{figure}[!tbp]
\centering
\includegraphics[width=1\textwidth]{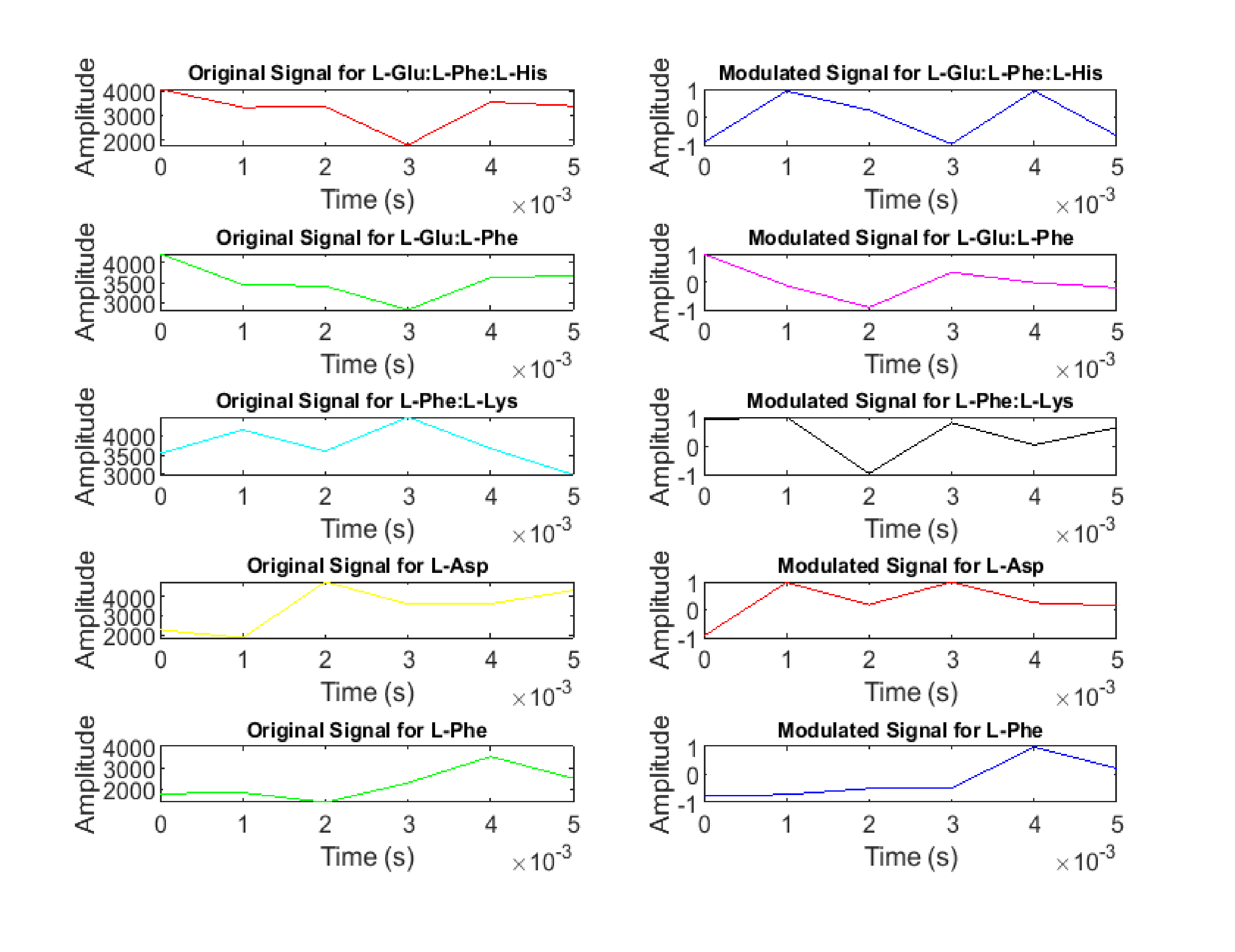}
\caption{The diagram displays the periods, measured in seconds, of proteinoids labelled as L-Glu:L-Phe:L-His, L-Glu:L-Phe, L-Phe:L-Lys, L-Asp, and L-Phe, after being exposed to both white and black light sources, as well as frequency modulated signals using a carrier signal of 200 Hz and a frequency deviation of 50 Hz.
}
\label{vsgffafavasb}
\end{figure}

The illustration (Fig.~\ref{vsgffafavasb}) depicts the principle of frequency modulation, a method of modifying the frequency of a carrier signal based on the message signal. Frequency modulation finds its application in various domains such as radio broadcasting, satellite television, audio synthesis and computing applications.

To avoid aliasing, a distortion of the signal caused by insufficient sampling, the sampling frequency must be at least twice the carrier frequency. The term for this is the Nyquist criterion~\cite{Frequency1442023,IIR,Butterworth}. The graphs suggest that there exists a positive correlation between the amplitude of the original signal and the frequency deviation of the modulated signal. Specifically, as the amplitude of the original signal increases, the frequency deviation of the modulated signal also increases. Additionally, it is worth noting that the modulated signal maintains a constant amplitude, which is in contrast to the original signal. Furthermore, the modulated signal exhibits a greater bandwidth than the original signal.

\begin{figure}[!tbp]
\centering
\includegraphics[width=0.8\textwidth]{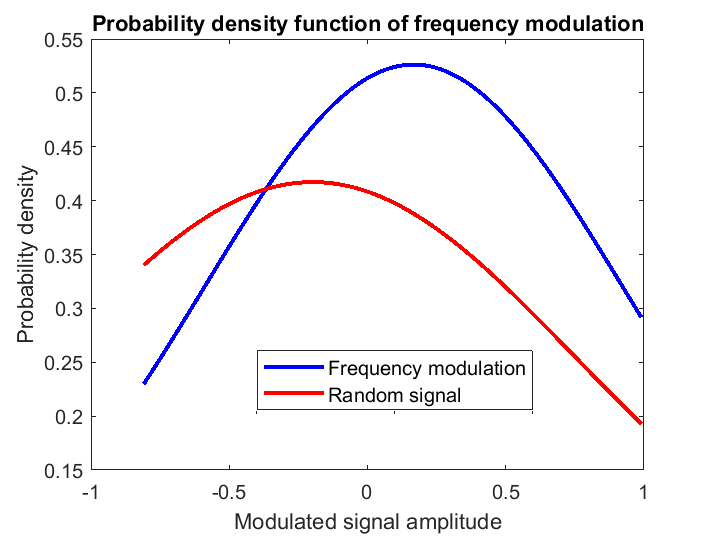}
\caption{This diagram illustrates the probability density function of the modulated signal amplitude.The findings indicate that the amplitude of the modulated signal exhibits a Gaussian distribution, with a majority of the values clustering around the mean and a smaller number of values at the tails. The previous claim implies that the amplitude of the signal remains relatively constant despite the application of frequency modulation, and any fluctuations in amplitude are both unpredictable and exhibit symmetry. The Gaussian distribution is a suitable approximation for this kind of data.
Here, $\mu$=-0.0008 and $\sigma$=0.7076. The findings indicate that the frequency modulation exhibits a comparable configuration to the random signal, although with different metrics for mean ($\mu$) and standard deviation ($\sigma$).  In contrast, the random signal is characterised by a sample mean of 0 and a sample standard deviation of 1. The application of frequency modulation results in a leftward shift of the distribution and a reduction in its width compared to that of the random signal. The Gaussian distribution is a suitable approximation for both signals~\cite{giron2017digital}. }
\label{fdsagsagfdsbgd}
\end{figure}

In systems or automatic control theory, Gaussian noise and perturbation models are commonly used. The standard probability density function (PDF) includes the following mathematical equation (Eq.~\ref{re}):
\begin{equation} \label{re}
fy(v) = \frac{e^{-(v-\mu)^2/2\sigma^2}}{\sqrt{2\pi}\sigma}
\end{equation}
The value of $\mu$ represents the average or central tendency of the random variable, while the value of $\sigma$ indicates the amount of variation or spread in the data. In Figure~\ref{fdsagsagfdsbgd}, there is a signal that has been generated randomly and follows a normal probability distribution function. The data was generated using MATLAB randn() function~\cite{giron2017digital}. The probability density function of frequency modulation can be seen in Figure~\ref{fdsagsagfdsbgd}.

For many years, scientists have been fascinated by the intriguing electrical oscillations exhibited by proteinoid microspheres~\cite{hsu1971conjugation,fox1976evolutionary,ponnamperuma1989experimental}. The way in which pendulum damped periodic motion and other physical phenomena relate to them has sparked a great deal of interest~\cite{kragel1994surface}.
In the next few paragraphs, we will delve into the fascinating topic of how proteinoid microspheres produce electrical oscillations and how this phenomenon is connected to pendulum damped periodic motion. To begin with, it's important to grasp the idea of electrical oscillations in proteinoid microspheres. Self-organising structures made up of amino acids have been discovered to generate electrical oscillations on their own. Voltage spikes are produced in the form of oscillations that can persist for a few minutes before gradually fading into a signal with low amplitude. Scientists believe that the electrical oscillations observed in the microsphere are a result of the interactions between its components~\cite{przybylski1985excitable}. These oscillations are thought to be connected to the microsphere's self-organisation.

What is the relationship between electrical oscillations and damped periodic motion of a pendulum? Both of these phenomena pertain to oscillatory systems wherein energy is gradually dissipated. However, the electrical oscillations observed in proteinoid microspheres arise from the interactions among the constituent amino acids.
What is the relationship between electrical oscillations of protenoids and chaotic systems?
Protein electrical oscillations exhibit a chaotic behaviour, which means that even a slight alteration in the system can result in significant and unforeseeable changes in the final outcome. The utilisation of this particular system has the potential to enhance our comprehension of intricate biological systems, while also enabling us to devise effective approaches for managing unstable states~\cite{malchow2002dynamical}. 

\begin{figure}[!tbp]
\centering
\includegraphics[width=1\textwidth]{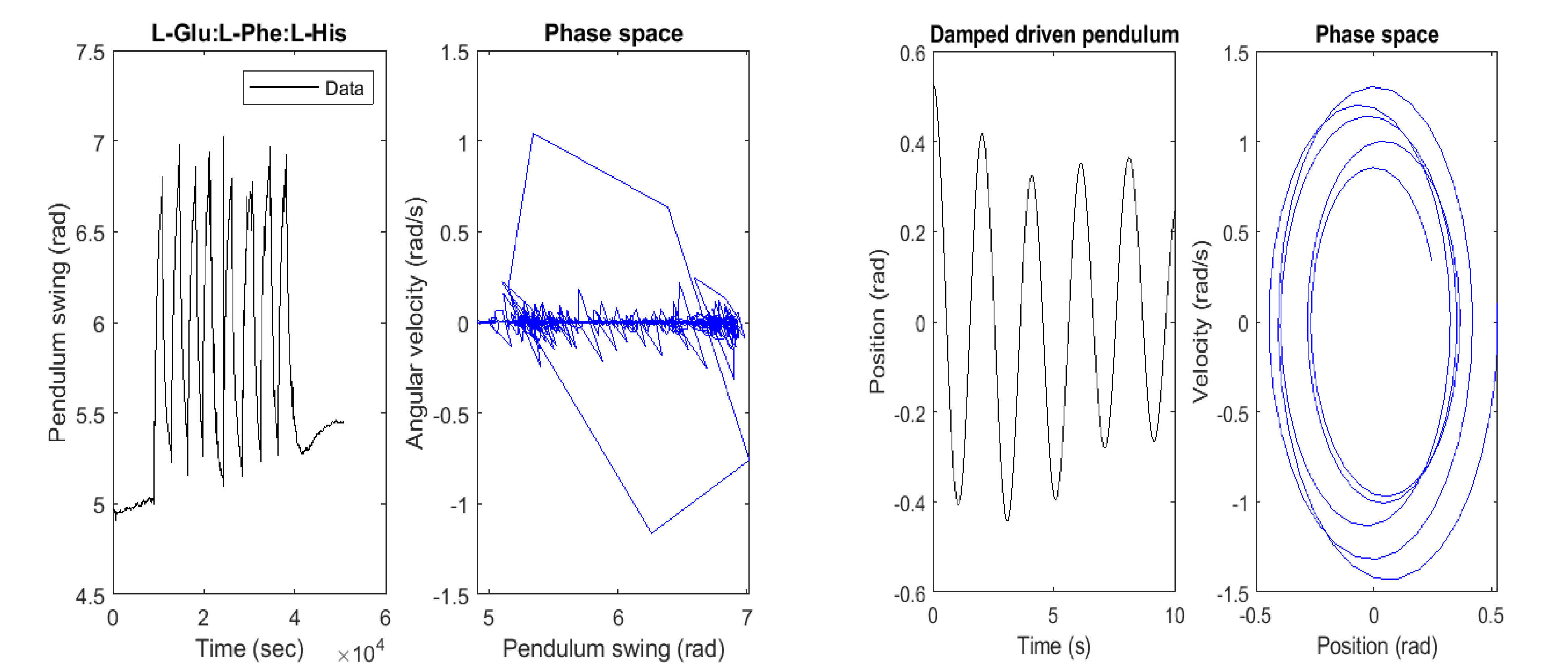}
\caption{The figure displays a correlation between the electrical oscillations of proteinoids and the periodic motion of a pendulum. Proteinoids have the ability to create basic electrical circuits, as evidenced by the oscillations visible in the image. This is a topic of research. The pendulum is a well-known illustration of periodic motion, as it swings back and forth in a foreseeable pattern. The aforementioned occurrences showcase the intricate conduct that may arise from basic structures.}
\label{afvsdbdbdfa}
\end{figure}

On the left-hand side (Fig.~\ref{afvsdbdbdfa}), there is a graph that displays the potential versus time data for the proteinoid L-Clu:L-Phe:L-His. The potential refers to the level of electrical activity exhibited by a neuron, specifically in the context of protenoids microspheres. It is revealed that the potential undergoes periodic oscillations that change in both amplitude and frequency. The microspheres' oscillations are occasionally interrupted by sharp spikes, which suggest that they are firing action potentials. On the right-hand side, you can see a plot that displays the relationship between the potential and its derivative in phase space. The concept of phase space refers to a theoretical space that illustrates the evolution of a system's state over time. It is an abstract representation that helps to visualise the changes that occur in a system as time progresses. In the plot, we can observe that there is a fascinating interplay between the potential and its derivative, which results in a spiral shape being formed in the phase space. When we observe a spiral shape in a system, it typically suggests that the system is nonlinear and chaotic~\cite{epstein1991nonlinear}. When there are spikes in the potential, it means that there are sudden jumps happening in the phase space. The key distinction between these plots lies in the fact that the earlier ones were generated using a basic pendulum model. This model is essentially a mechanical system that adheres to a differential equation. In the earlier plots, we can observe the position and velocity of the pendulum over time, as well as in the phase space. The way in which the pendulum moved was related to its position and velocity in a predictable way known as simple harmonic motion. This relationship between the two variables formed a shape that resembled an ellipse when plotted in what is known as phase space. The presence of an ellipse shape in the system suggests that it is both linear and periodic. In the proteinoid plots, there were sudden spikes or jumps observed in the phase space~\cite{vistnes2018physics}.



\section{Conclusion}

In conclusion, the spiking frequency modulation of proteinoids with light is a fascinating concept that can open up a range of possibilities for increased efficiency in unconventional computing, synthetic biology and bio-engineering. Despite the complexities and challenges that still remain in fully understanding the range of effects and applications of this phenomenon, the potential for a revolution in the computing world is undeniable.

\section*{Acknowledgement}

The research was supported by EPSRC Grant EP/W010887/1 ``Computing with proteinoids''. Authors are grateful to David Paton for helping with SEM imaging and to Neil Phillips for helping with instruments.



\end{document}